\documentclass[12pt, twoside, epsf]{article}

\usepackage{color,graphicx,times}
\usepackage{amsmath, amssymb, graphics}
\usepackage{latexsym}

\newcommand{\mathsym}[1]{{}}
\newcommand{\unicode}[1]{{}}

\def\Set#1{ \{#1\}}

\makeatletter
\long\def\M#1{\leavevmode\setbox\@tempboxa\hbox{#1}\@tempdima\fboxrule
    \advance\@tempdima \fboxsep \advance\@tempdima \dp\@tempboxa
   \hbox{\lower \@tempdima\hbox
  {\vbox{\hrule \@height \fboxrule
          \hbox{  \hskip\fboxsep
          \vbox{\vskip\fboxsep \box\@tempboxa\vskip\fboxsep}\hskip
                 \fboxsep\vrule \@width \fboxrule}%
                  }}}}
\makeatother

\let \ttorg \tt \def \tt{\ttorg \obeyspaces}

\begin{document}

\date{}

\title{\bf Non-Commutative Worlds  and Classical Constraints}

\author{Louis H. Kauffman \\
  Department of Mathematics, Statistics and Computer Science \\
  University of Illinois at Chicago \\
  851 South Morgan Street\\
  Chicago, IL, 60607-7045}

\maketitle
  
\thispagestyle{empty}

\abstract{This paper reviews results about discrete physics and non-commutative worlds and explores further the structure and consequences of constraints linking classical calculus and discrete 
calculus formulated via commutators. In particular we review how the formalism of generalized non-commutative electromagnetism follows from a first order constraint and how, via the Kilmister equation,
relationships with general relativity follow from a second order constraint. It is remarkable that a second order constraint, based on interlacing the commutative and non-commutative worlds, leads to an equivalent 
tensor equation at the pole of geodesic coordinates for general relativity.}\\


\section{Introduction}
Aspects of gauge
theory, Hamiltonian mechanics, relativity and quantum mechanics arise naturally in the mathematics of a non-commutative
framework for calculus and differential geometry. 
In this paper, we give a review of our previous
results about discrete physics and non-commutative worlds and an introduction to recent work of the author and Anthony Deakin \cite{KauffDeak}. 
In examining the foundations of that work we find new points of view and clarity of proofs as expressed in the later sections of this paper. 
A key feature of the present paper is a new and concise derivation of the second constraint in Section 4 and a detailed derivation of the related Kilmister equation in Section 5. In Section 6 we determine the 
third constraint by similar means. At this time, physics associated with the higher order constraints is not known.\\

We begin by examining discrete dynamical systems. In our exposition the  simplest discrete system corresponds to the square root of minus one, seen as an oscillation between one and minus one. 
This way thinking about $i$ as an {\em iterant} is explained below. By starting with a discrete time series of positions, one has immediately a non-commutativity of observations, since the measurement of velocity involves the tick of the clock and the measurement of position does not demand the tick of the clock. Commutators that arise from discrete
observation suggest a non-commutative calculus, and this calculus leads to a generalization of 
standard advanced calculus in terms of a non-commutative world. In a non-commutative world,
all derivatives are represented by commutators. We review how non-commutative worlds are related to quantum physics and classical physics  and review our version of the Feynman-Dyson derivation of the formalism of electromagnetic gauge theory. The rest of the paper then investigates algebraic constraints that bind the commutative and non-commutative worlds. These constraints are demands that time derivatives behave in the non-commutative world in analogy to their counterparts in standard advanced calculus. It is one constraint of this type that gives rise to our version of the Feynman-Dyson derivation of electromagnetic formalism. The standard first order constraint requires a quadratic Hamiltonian and so begins a story showing how classical physics arises mathematically from the constraints. The second order constraint turns out, remarkably,  to be equivalent
to a tensor equation at the pole of canonical coordinates in a relativistic framework. We call this tensor equation the {\it Kilmister equation} and it is studied in Section 5 and in our paper with Deakin \cite{KauffDeak}.
\bigbreak

Section 2 is a self-contained version of the concepts in this paper, starting with the non-commutativity of discrete measurements, the introduction of time-shifting operators and  
the square root of minus one seen as a discrete oscillation, a clock. We proceed from there and analyze the position of the square root of minus one in relation to discrete systems and quantum mechanics.
We end this section by fitting together these observations into the structure of the Heisenberg commutator $$[p,q] = i \hbar.$$ 
Section 3 is a review of the context of non-commutative worlds with discussion of the Feynman-Dyson derivation. This section generalizes the concepts in Section 2 and places them in the wider context of non-commutative worlds. The key to this generalization is our method of embedding discrete
calculus in the non-commutative context.
Section 4 discusses constraints on non-commutative worlds that are imposed by asking for correspondences between
forms of classical differentiation and the derivatives represented by commutators in a correpondent non-commutative world.
This discussion of constraints parallels work of Tony Deakin \cite{Deakin1,Deakin2} and is continued in joint work of the author and Deakin \cite{KauffDeak}. 
At the level of the second constraint we encounter issues related to general relativity and find that at the pole of a canonical system of coordinates the second order constraint is equivalent to the Kilmister equation
$$K_{ab} = g^{ef}(R_{ab;ef} + \frac{2}{3} R_{ae}R_{fb})=0$$
where  $a,b,e,f = 1,2, \cdots 4$ and $R$ is the curvature tensor corresponding to the metric $g_{ab}$ on spacetime. Section 5 gives a derivation of the Kilmister equation and its relation to the second order constraint,
following the original observations of Kilmister \cite{Kilmister}. In the present paper we give a proof that the second order constraint is equivalent to the Kilmister equation.  One can regard the Kilmister equation $K_{ab}=0$ as a higher order replacement for the vacuum Einstein equation $R_{ab}= 0$ (the vanishing of the Ricci tensor).
In \cite{KauffDeak} this approach to modifying general relativity, and some of its consequences are explored in detail.\\

Section 5 continues the constraints discussion in Section 4, showing how to generalize to higher-order constraints and obtains a commutator formula for the third order constraint. 
The appendix, Section 7,  is a very condensed review of the relationship of the Bianchi identity in differential geometry and the Einstein equations for general relativity. We then observe that every derivation in a non-commutative world comes equipped with its own Bianchi identity. This observation suggests another way  to investigate general relativity in the non-commutative context.
\bigbreak

\section{Time Series and Discrete Physics}

Consider elementary discrete physics in one dimension. Consider a time series of positions
$$x(t):  t = 0, \Delta t, 2\Delta t, 3\Delta t, \cdots . $$
We can define the velocity $v(t)$ by the formula 
$$v(t) = (x(t+ \Delta t) - x(t))/\Delta t = Dx(t)$$ 
where $D$ denotes this discrete derivative. In order to obtain $v(t)$ we need at least one tick  $\Delta t$ of the discrete clock.  We define a time-shift operator to handle the fact that once we have observed $v(t),$ the time has moved up by one tick. 
\bigbreak

\noindent{\bf We adjust the discrete derivative.}
We shall add an operator J that in this context accomplishes the time shift:
$$x(t)J = Jx(t+\Delta t).$$
We then redefine the derivative to include this shift:
$$Dx(t) = J(x(t+ \Delta t) - x(t))/\Delta t .$$
This readjustment of the derivative rewrites it so that the temporal properties of successive
observations are handled automatically.
\smallbreak

\noindent{\bf Discrete observations do not commute.}
Let $A$ and $B$ denote quantities that we wish to observe in the discrete system.
Let $AB$ denote the result of first observing $B$ and then observing $A.$
The result of this definition is that a successive observation 
of the form $x(Dx)$ is distinct from an observation of the form
$(Dx)x.$ In the first case, we first observe the velocity at time $t$, and then $x$ is measured at $t + \Delta t $. In the second case, we measure $x$ at $t $ and then 
measure the velocity.
\smallbreak

We measure the difference between these two results by taking a commutator $$[A,B] = AB - BA$$ and we get the following computations where we write  $\Delta x = x(t+ \Delta t) - x(t).$
$$x(Dx) = x(t)J( x(t+ \Delta t) - x(t))/\Delta t = Jx(t+ \Delta t)( x(t+ \Delta t) - x(t))/\Delta t.$$
$$(Dx)x = J(x(t+ \Delta t) - x(t))x(t)/\Delta t.$$
$$[x,Dx] = x(Dx) - (Dx)x  = (J/\Delta t)(x(t+ \Delta t) - x(t))^2  = J (\Delta x)^2/\Delta t$$
This final result is worth recording:
$$[x,Dx] = J ( \Delta x)^2/\Delta t.$$
From this result we see that the commutator of $x$ and$ Dx$ will be constant if $(\Delta x)^2/\Delta t = k$ is a constant. For a given time-step,  this
means that $$(\Delta x)^2 = k \Delta t$$  so that  $$\Delta x =  \pm \sqrt{(k \Delta t )}$$This is a Brownian process with diffusion constant equal to $k.$
\smallbreak

Thus we arrive at the result that any discrete process viewed in this framework of discrete observation has the basic commutator $$[x,Dx] = J ( \Delta x)^2/\Delta t, $$ generalizing a Brownian process and 
containing the factor $ ( \Delta x)^2/\Delta t $ that corresponds to the classical diffusion constant.
It is worth noting that the adjusment that we have made to the discrete derivative makes it into a commutator as follows:
$$Dx(t) = J(x(t+ \Delta t) - x(t))/\Delta t  = (x(t)J - Jx(t))\Delta t = [x(t), J]/\Delta t.$$
By replacing discrete derivatives by commutators we can express discrete physics in many variables in a context of non-commutative algebra. We enter this generalization in the next section of the paper.
\smallbreak

A simplest and fundamental instance of these ideas is seen in the structure of $i  = \sqrt{-1}.$ We view $i$ as an {\em iterant} \cite{Iterants},
a discrete elementary dynamical system repeating in time the values $\{\cdots -1,+1,-1,+1, \cdots\}.$ One can think of this system as 
resulting from the attempt to solve $i^2 = -1$ in the form $i = -1/i.$ Then one iterates the transformation $ x \longrightarrow -1/x$ and finds the
oscillation from a starting value of $+1$ or $-1.$ In this sense $i$ is identical in concept to a {\em primordial time.} Furthermore the algebraic structure
of the complex numbers emerges from two conjugate views of this discrete series as $[-1,+1]$ and $[+1,-1]$. We introduce a temporal shift operator
$\eta$ such that $\eta [-1, +1] = [+1,-1]\eta$ and $\eta^2 = 1$ (sufficient to this purpose). Then we can define $i = [-1,+1]\eta,$ endowing it
with one view of the discrete oscillation and the sensitivity to shift the clock when interacting with itself or with another operator.  
Note that if $e=[-1,+1]$ and we take $[a,b][c,d] = [ab,cd]$ and $-[a,b] = [-a,-b],$ then $$e^2 = \eta^2 = 1$$ and $$e \eta + \eta e = 0.$$ Hence, with 
$$i = e \eta$$ we have $$i^2 = e \eta e \eta = -e^2 \eta^2 = -1.$$

Here we see $i$ emerge in the non-commutative context of the Clifford algebra generated by $e$ and $\eta,$ and we see that
in this way, $i$ becomes inextricably 
identified with elemental time, and so the physical substitution of $i t$ for $t$ (Wick rotation) becomes, in this epistemology, an 
act of recognition of the nature of time. One does not have an increment of time all alone as in classical $t. $
One has $it,$ a combination of an interval and the elemental dynamic that is time. 
With this understanding, we can return to the commutator for a discrete process and use $i \Delta t$ for the temporal increment.
\smallbreak

We found that discrete observation led to the commutator equation
$$[x,Dx] = J (\Delta x)^2/ \Delta t$$
which we will simplify to
$$[q, p/m] = (\Delta x)^2/\Delta t.$$
taking $q$ for the position $x$ and $p/m$  for velocity, the time derivative of position and ignoring the
time shifting operator on the right hand side of the equation.
\smallbreak

Understanding that $\Delta t$ should be replaced by $ i \Delta t,$  and that, by comparison with the physics of a process at the Planck scale one can take
$$ (\Delta x)^2/\Delta t = \hbar /m, $$
we have
$$[q, p/m] = (\Delta x)^2/i \Delta t = -i \hbar /m,$$
whence
$$[p,q] = i\hbar,$$
and we have arrived at Heisenberg's fundamental relationship between position and momentum. This mode of arrival is predicated on the recognition that $i \Delta t$ represents an interactive interval of time.
In the notion of time there is an inherent clock and an inherent shift of phase that enables a synchrony, a precise dynamic beneath the apparent dynamic of the observed process. 
Once this substitution is made, once the imaginary value is placed in the temporal circuit, the patterns of quantum mechanics appear. 
In this way, quantum mechanics can be seen to emerge from the discrete.
\smallbreak

\section{Review of Non-Commutative Worlds}
Now we begin a general introduction to non-commutative worlds and to a non-commutative discrete calculus.
Our approach begins in an algebraic framework that naturally contains the formalism of the
calculus, but not its notions of limits or constructions of spaces with specific locations, points and trajectories. Many patterns
of  physical law fit well into such an abstract framework. In this viewpoint one dispenses with continuum
spacetime and replaces it by algebraic structure. Behind that structure, space stands ready to be constructed, by discrete
derivatives and patterns of steps, or by starting with a discrete pattern in the form of a diagram, a network, a lattice, a knot, or a
simplicial complex, and elaborating that structure until the specificity of spatio-temporal locations appear.
\bigbreak 

Poisson brackets allow one to connect
classical notions of location with the non-commutative algebra used herein.  Below the level of the Poisson brackets is a treatment of 
processes and operators as though they were variables in the same context as the variables in the classical calculus.
In different degrees one lets go of the notion of classical variables and yet retains their form, as one makes a descent into the 
discrete. The discrete world of non-commutative operators is a world linked to our familiar world of continuous and commutative
variables. This linkage is traditionally exploited in quantum mechanics to make the transition from the classical to the quantum.
One can make the journey in the other direction, from the discrete and non-commutative to the ``classical" and commutative, but 
that journey requires powers of invention and ingenuity that are the subject of this exploration. It is our conviction that the 
world is basically simple. To find simplicity in the complex requires special attention and care.
\bigbreak

In starting from a discrete point of view one thinks of a sequence of states of the world $S,S',S'',S''', \cdots$ where $S'$
denotes the state succeeding $S$ in discrete time. It is natural to suppose that there is some measure of difference
$DS^{(n)} = S^{(n+1)} - S^{(n)},$ and some way that states $S$ and $T$ might be combined to form a new state $ST.$ We can thus 
think of world-states as operators in a non-commutative algebra with a temoporal derivative $DS = S' - S.$ At this bare level of 
the formalism the derivative does not satisfy the Leibniz rule. In fact it is easy to verify that 
$D(ST) = D(S)T + S'D(T).$ Remarkably, the Leibniz rule, and hence the formalisms of Newtonian calculus can be restored with the
addition of one more operator $J.$ In this instance $J$ is a temporal shift operator with the property that
$SJ = JS'$ for any state $S.$ We then see that if $\nabla S = JD(S) = J(S'-S).$ then $\nabla(ST) = \nabla(S)T + S\nabla(T)$ for any 
states $S$ and $T.$ In fact $\nabla(S) = JS' - JS = SJ - JS = [S,J],$ so that this adjusted derivative is a commutator in the general
calculus of states. This, in a nutshell, is our approach to non-commutative worlds. We begin with a very general framework that is
a non-numerical calculus of states and operators. It is then fascinating and a topic of research to see how physics and mathematics
fit into the frameworks so constructed.
\bigbreak

Constructions are performed in a Lie algebra $\cal A.$   
One may take $\cal A$ to be a specific matrix Lie algebra, or abstract Lie algebra.
If $\cal A$ is taken to be an abstract Lie algebra, then it is convenient to use the universal
enveloping algebra so that the Lie product can be expressed as a commutator. In making general constructions of
operators satisfying certain relations, it is understood that one can always begin with a free algebra and make
a quotient algebra where the relations are satisfied.
\bigbreak

On $\cal A,$ a variant of calculus  is built by
defining derivations as commutators (or more generally as Lie products). For a fixed $N$ in $\cal A$ one defines
$$\nabla_N : \cal A \longrightarrow \cal A$$ by the formula
$$\nabla_{N} F = [F,N] = FN - NF.$$
$\nabla_N$ is a derivation satisfying the Leibniz rule.  
$$\nabla_{N}(FG) = \nabla_{N}(F)G + F\nabla_{N}(G).$$
\bigbreak

\noindent {\bf Discrete Derivatives are Replaced by Commutators.} 
There are many motivations for replacing derivatives by commutators. If
$f(x)$ denotes (say) a function of a real variable
$x,$ and $\tilde{f}(x) = f(x+h)$ for
a fixed increment $h,$ define the {\em discrete derivative} $Df$ by the formula $Df = (\tilde{f} - f)/h,$ and find that
the Leibniz rule is not satisfied. One has the basic formula for the discrete derivative
of a product: $$D(fg) = D(f)g + \tilde{f}D(g).$$
Correct this deviation from the Leibniz rule by introducing a new non-commutative operator $J$ with the property that 
$$fJ = J\tilde{f}.$$ Define a new discrete derivative in an extended non-commutative algebra by the formula
$$\nabla(f) = JD(f).$$ It follows at once that 
$$\nabla(fg) = JD(f)g + J\tilde{f}D(g) = JD(f)g + fJD(g) = \nabla(f)g + f\nabla(g).$$
Note that $$\nabla(f) = (J\tilde{f} - Jf)/h =
(fJ-Jf)/h = [f,J/h].$$ In the extended algebra, discrete derivatives are represented by commutators, and satisfy the Leibniz rule. 
One can regard discrete calculus as a subset of non-commutative
calculus based on commutators.
\bigbreak

\noindent {\bf Advanced Calculus and Hamiltonian Mechanics or Quantum Mechanics in a Non-Commutative World.} In $\cal A$ there
are as many derivations as there are elements of the algebra, and these derivations behave quite wildly with respect to one another. If
one takes the concept of {\em curvature} as the non-commutation of
derivations, then $\cal A$ is a highly curved world indeed. Within $\cal A$ one can build 
a tame world of derivations that mimics the
behaviour of flat coordinates in Euclidean space. The description of the
structure of $\cal A$ with respect to these flat coordinates contains many of the
equations and patterns of mathematical physics.
\bigbreak

\noindent The
flat coordinates $Q^i$ satisfy the equations below with the $P_j$ chosen to represent differentiation with
respect to $Q^j$:

$$[Q^{i}, Q^{j}] = 0$$
$$[P^{i},P^{j}]=0$$
$$[Q^{i},P^{j}] = \delta^{ij}.$$
Here $\delta^{ij}$ is the Kronecker delta, equal to $1$ when $i = j$ and equal to $0$ otherwise.
Derivatives are represented by commutators. 
$$\partial_{i}F = \partial F/\partial Q^{i} = [F, P^{i}],$$
$$\hat{\partial_{i}}F = \partial F/\partial P^{i} = [Q^{i},F].$$ 
Our choice of commutators guarantees that the derivative of a variable with respect to itself is one and
that the derivative of a variable with respect to a distinct variable is zero. Furthermore, the commuting of 
the variables with one another guarantees that mixed partial derivatives are independent of the order of differentiation. This is a flat non-commutative world.
\bigbreak

Temporal derivative is represented by commutation with a special (Hamiltonian) element $H$ of the algebra:
$$dF/dt = [F, H].$$
(For quantum mechanics, take $i\hbar dA/dt = [A, H].$)
These non-commutative coordinates are the simplest flat set of
coordinates for description of temporal phenomena in a non-commutative world.
\bigbreak

\noindent {\bf Hamilton's Equations are Part of the Mathematical Structure of Non-Commutative Advanced Calculus.}
$$dP^{i}/dt = [P^{i}, H] = -[H, P^{i}] = -\partial H/\partial Q^{i}$$
$$dQ^{i}/dt = [Q^{i}, H] = \partial H/\partial P^{i}.$$ 
These are exactly Hamilton's equations of motion. The pattern of
Hamilton's equations is built into the system.
\bigbreak

\noindent {\bf The Simplest Time Series Leads to the Diffusion Constant and Heisenberg's Commuator.}
Consider a time series $\{Q,Q',Q'',...\}$ with commuting scalar values.
Let $$\dot{Q} = \nabla Q = JDQ = J(Q'-Q)/\tau$$ where $\tau$ is an elementary time step (If $Q$ denotes a times series value at time
$t$, then 
$Q'$ denotes the value of the series at time $t + \tau.$). The shift operator $J$ is defined by the equation
$QJ = JQ'$
where this refers to any point in the time series so that $Q^{(n)}J = JQ^{(n+1)}$ for any non-negative integer $n.$
Moving $J$ across a variable from left to right, corresponds to one tick of the clock. This discrete,
non-commutative time derivative satisfies the Leibniz rule. 
\bigbreak

This derivative $\nabla$ also fits a significant pattern of discrete observation. Consider the act of observing $Q$ at a given time
and the act of observing (or obtaining) $DQ$ at a given time. 
Since $Q$ and $Q'$ are ingredients in computing $(Q'-Q)/\tau,$ the numerical value associated with $DQ,$ it is necessary  to let the
clock tick once, Thus, if one first observe
$Q$ and then obtains $DQ,$ the result is different (for the $Q$ measurement) if one first obtains $DQ,$ and then observes $Q.$ In the
second case, one finds the value $Q'$ instead of the value $Q,$ due to the tick of the clock. 
\bigbreak

\begin{enumerate}
\item Let $\dot{Q}Q$ denote the sequence: observe $Q$, then obtain $\dot{Q}.$ 
\item Let $Q\dot{Q}$ denote the sequence: obtain $\dot{Q}$, then observe $Q.$ 
\end{enumerate}
\bigbreak

The commutator $[Q, \dot{Q}]$ expresses the difference between these two orders of discrete measurement.
In the simplest case, where the elements of the time series are commuting scalars, one has
$$[Q,\dot{Q}] = Q\dot{Q} - \dot{Q}Q =J(Q'-Q)^{2}/\tau.$$
Thus one can interpret the equation $$[Q,\dot{Q}] = Jk$$ ($k$ a constant scalar) as $$(Q'-Q)^{2}/\tau = k.$$ This means
that the process is a walk with spatial step $$\Delta = \pm \sqrt{k\tau}$$ where $k$ is a constant. In other words, one 
has the equation
$$k = \Delta^{2}/\tau.$$ 
This is the diffusion constant for a Brownian walk.
A walk with spatial step size  $\Delta$ and time step $\tau$ will satisfy the commutator equation above
exactly when the square of the spatial step divided by the time step remains constant. This
shows that the diffusion constant of a Brownian process is a structural property of that process, independent of considerations of
probability and continuum limits.  
\bigbreak

Thus we can write (ignoring the timeshift operator $J$) $$[Q,\dot{Q}] = (\Delta Q)^2/\tau.$$
If we work with physics at the Planck scale, then we can take $\tau$ as the Planck time and $\Delta Q$ as the Planck length.
Then $$(\Delta Q)^2 / \tau = \hbar/m$$ where $m$ is the Planck mass. However, we shall also Wick rotate the time from $\tau$ to
$i \tau$ justifying $i \tau$ on the principle (described above) that $\tau$ should be multiplied by $i$ to bring time into coincidence with an
elemental time that is both a temporal operator ($i$) and a value ($t$). With this we obtain $$[Q,\dot{Q}]  =- i \hbar/m$$ or
$$[m \dot{Q},Q]  = i \hbar,$$ and taking $P = m\dot{Q},$ we have finally $$[P,Q]  = i \hbar.$$ Heisenberg's commutator for quantum mechanics is
seen in the nexus of discrete physics and imaginary time.
\bigbreak

\noindent {\bf Schroedinger's Equation is Discrete.}
Here is how the Heisenberg form of Schroedinger's equation fits in this context. Let $J=(1 - \frac{i}{\hbar} H \Delta t).$
Then $\nabla \psi = [\psi, J/\Delta t],$ and we calculate
$$\nabla \psi = \frac{1}{\Delta t} [\psi J - J \psi]$$
$$= \psi[(1 -  \frac{i}{\hbar}  H \Delta t)/\Delta t] - [(1 -  \frac{i}{\hbar}  H \Delta t)/\Delta t] \psi =  - \frac{i}{\hbar}[\psi, H].$$
Thus
$$\nabla \psi = - \frac{i}{\hbar}[\psi, H].$$
This is exactly the form of the Heisenberg equation.\\

Another way to think about this operator $J = (1 - \frac{i}{\hbar} H \Delta t)$ is as an approximation to $e^{- \frac{i}{\hbar} H \Delta t}.$
We can then see our discrete model behaving {\it exactly} in the framework of a calculus using {\it square zero infinitesimals} \cite{Bell}.
Let us recall the bare bones of this model for calculus. We utilize an algebraic entity denoted here by $dt$ such that $(dt)^2 = 0$ and an extended real number system 
$$R^{\sharp} = \{a + b dt \}$$ where it is understood that $a$ and $b$ are standard real numbers and that $a + b dt = a' + b' dt$ if and only if $a=a'$ and $b=b'.$ It is given that 
$dt > 0$ and $dt < r$ for any positive real number $r.$ We multipy by assuming distributivity and using the nilpotence of $dt.$ Thus
$$(a + b dt) (e + f dt) = ae + (af + be) dt.$$ The special infinitesimal $dt$ is not invertible, but for those functions that have a well-defined extension to $R^{\sharp}$ we can define the derivative by the formula
$$F(t + dt) = F(t) + \dot{F}(t)dt.$$ In the case of the exponential function, we have $$e^{r + s dt} = e^{r}(1 + s dt)$$ as the {\it definition} of this extension of the exponential function. The reader should note that
this means that $$e^{s dt} = 1 + s dt$$ and that this is exactly the result obtained by substitution into the power series $$e^{x} = 1 + x + x^2 /2! + x^3 /3! + \cdots$$ and using the nilpotency of the $dt.$ 
Thus we find that $e^{a(t + dt)} = e^{at}(1 + adt) = e^{at} + ae^{at}dt$ and therefore the derivative of $e^{at}$ with respect to $t$ is $a e^{at},$ as expected. In the same vein, 
$$e^{idt} = 1 + i dt$$ and 
$$e^{i dt} = cos(dt) + i sin(dt),$$ from which we conclude that $$cos(dt) = 1$$ and $$sin(dt) = 0.$$ With this rapid course in infinitesimal calculus we return to time shifter $J.$\\

In the nilpotent infinitesimal calculus we have $$J =  (1 - \frac{i}{\hbar} H dt) = e^{- \frac{i}{\hbar} H dt}.$$ and 
$$J^{-1} = (1 + \frac{i}{\hbar} H dt) = e^{+\frac{i}{\hbar} H dt}.$$ 
Note that we can formally multiply $(1 - \frac{i}{\hbar} H dt)((1 + \frac{i}{\hbar} H dt)$ and obtain $1$ since $(dt)^2 = 0$.
We continue to think of $dt$ as a discrete increment, even though it is infinitesimal.
Our time-shift formula is 
$$J \psi(t + dt) = \psi(t) J$$ or, equivalently,
$$\psi(t + dt) = J^{-1} \psi(t) J.$$\\

With this in mind we calculate and find:
$$\psi(t + dt) = (1 + \frac{i}{\hbar} H dt) \psi(t) (1 - \frac{i}{\hbar} H dt) = (\psi(t) +  \frac{i}{\hbar}dt H \psi(t)) ( (1 - \frac{i}{\hbar} dt H)$$
$$ = \psi(t) +  \frac{i}{\hbar}dt (H \psi(t) - \psi(t) H) = \psi(t) - \frac{i}{\hbar}[\psi, H] dt$$
Thus 
$$\psi(t + dt) = \psi(t) - \frac{i}{\hbar}[\psi, H] dt$$
from which we conclude that
$$\dot{\psi} = - \frac{i}{\hbar}[\psi, H] ,$$
arriving again at the Heisenberg version of Schroedinger's equation in the context of nilpotent calculus.\\

\noindent {\bf Dynamical Equations Generalize Gauge Theory and Curvature.} 
One can take the general dynamical
equation in the form 
$$dQ^{i}/dt = {\cal G}_{i}$$ where $\{ {\cal G}_{1},\cdots, {\cal G}_{d} \}$
is a collection of elements of $\cal A.$ Write ${\cal G}_{i}$
relative to the flat coordinates via ${\cal G}_{i} = P_{i} -  A_{i}.$
This is a definition of $A_{i}$ and $\partial F/\partial Q^{i} = [F,P_{i}].$ The formalism of gauge theory appears
naturally. In particular, if $$\nabla_{i}(F) = [F, {\cal G}_{i}],$$ then one has
the curvature $$[\nabla_{i}, \nabla_{j}]F = [R_{ij}, F]$$
and 
$$R_{ij} = \partial_{i} A_{j} - \partial_{j} A_{i} + [A_{i}, A_{j}].$$  This is the well-known formula for the curvature of a gauge
connection. Aspects of geometry arise naturally in this context, including the Levi-Civita
connection (which is seen as a consequence of the Jacobi identity in an appropriate non-commutative world).  
\bigbreak

One can consider the consequences of the commutator $[Q^{i}, \dot{Q^{j}}] = g^{ij}$,
deriving that  
$$\ddot{Q^{r}} = G_{r} + F_{rs}\dot{Q^{s}} + \Gamma_{rst}\dot{Q^{s}}\dot{Q^{t}},$$
where $G_{r}$ is the analogue of a scalar field, $F_{rs}$ is the analogue of a gauge field and $\Gamma_{rst}$ is the Levi-Civita
connection associated with $g^{ij}.$
This decompositon of the acceleration is uniquely determined by the given framework \cite{NCW,Glafka,Diff}
\bigbreak

\noindent {\bf Non-commutative Electromagnetism and Gauge Theory.}
One can use this context to revisit the Feynman-Dyson derivation \cite{Dyson} of electromagnetism from commutator equations, 
showing that most of the derivation is independent of any choice of commutators, but highly dependent upon the choice of definitions
of the derivatives involved. Without any assumptions about initial commutator equations, but taking the right (in some sense simplest)
definitions of the derivatives one obtains a significant generalization of the result of Feynman-Dyson.  We give this derivation in \cite{KN:QEM} and in \cite{NCW,Glafka,Diff}
using diagrammatic algebra to clarify the structure. In this section we use $X$ to denote the
position vector rather than $Q,$ as above, and the partial derivatives $\{ \partial_{1}, \partial_{2}, \partial_{3} \}$ are each covariant derivatives represented by commutators with 
$\{\dot{X_{1}},\dot{X_{2}},\dot{X_{2}}\}$ respectively.
\bigbreak

\noindent {\bf Theorem} With the appropriate [see below] definitions of the operators, and taking
$$\nabla^{2} = \partial_{1}^{2} + \partial_{2}^{2} + \partial_{3}^{2}, \,\,\, B = \dot{X} \times \dot{X} \,\,\, \mbox{and} \,\,\, E =
\partial_{t}\dot{X}, \,\,\, \mbox{one has}$$

\begin{enumerate}
\item $\ddot{X} = E + \dot{X} \times B$
\item $\nabla \bullet B = 0$
\item $\partial_{t}B + \nabla \times E = B \times B$
\item $\partial_{t}E - \nabla \times B = (\partial_{t}^{2} - \nabla^{2})\dot{X}$
\end{enumerate}
\bigbreak

The key to the proof of this Theorem is the definition of the time derivative. This definition is as follows
$$\partial_{t}F = \dot{F} - \Sigma_{i}\dot{X_{i}}\partial_{i}(F) =  \dot{F} - \Sigma_{i} \dot{X_{i}}[F, \dot{X_{i}}]$$
for all elements or vectors of elements $F.$  The definition creates a
distinction between space and time in the non-commutative world. It can be regarded as an articulation of one extra constraint of the first order in the sense that 
we describe in the next section, Section 4, of this paper.\\

A calculation reveals that
$$\ddot{X} = \partial_{t}\dot{X} + \dot{X} \times (\dot{X} \times \dot{X}).$$
This suggests taking $E = \partial_{t}\dot{X}$ as the electric field, and $B = \dot{X} \times \dot{X}$
as the magnetic field so that the Lorentz force law 
$$\ddot{X} = E + \dot{X} \times B$$
is satisfied.
\bigbreak

\noindent 
This result can be applied to produce many discrete models of the Theorem. These models show that, just as the commutator $[X, \dot{X}] =
Jk$ describes Brownian motion in one dimension, a generalization of electromagnetism describes the interaction of triples of time
series in three dimensions.
\bigbreak

Taking $\partial_{t}F = \dot{F} - \Sigma_{i}\dot{X_{i}}\partial_{i}(F) =  \dot{F} - \Sigma_{i} \dot{X_{i}}[F, \dot{X_{i}}]$ as a definition
of the partial derivative with respect to time is a natural move in this context because there is  {\em no time variable $t$} in this 
non-commutative world. A formal move of this kind, matching a pattern from the commutative world to the mathematics of the non-commuative world
is the theme of the next section of this paper. In that section we consider the well known way to associate an operator to a product of commutative
variables by taking a sum over all permutations of products of the operators corresponding to the individual variables. This provides a way to 
associate operator expressions with expressions in the commuative algebra, and hence to let a classical world correspond or map to a non-commutative world. To bind these worlds more closely, we can ask that the formulas for taking derivatives in the commutative world should have symmetrized operator
product correspondences in the non-commutative world. In Sections 4 and 5 we show how the resulting {\it constraints}  are related to having a quadratic 
Hamiltonian (first order constraint) and to having a version of general relativity \cite{Deakin1,Deakin2, KauffDeak} (second order constraint). Such constraints can
be carried to all orders of derivatives, but the algebra of such constraints is, at the present time, in a very primitive state. We discuss some of the 
complexities of the constraint algebra in Section 6 of this paper.
\bigbreak

\noindent {\bf Remark.} While there is a large
literature on non-commutative geometry, emanating from the idea of replacing a space by its ring of  functions, work discussed herein
is not written in that tradition. Non-commutative geometry does occur here, in the sense of geometry occuring in the context of
non-commutative algebra. Derivations are represented by commutators. There are relationships between the present work and the
traditional non-commutative geometry, but that is a subject for further exploration. In no way is this paper intended to be an
introduction to that subject. The present summary is based on
\cite{Para,Kauff:KP,KN:QEM,KN:Dirac,Twist,NonCom,ST,Aspects,Boundaries,NCW,Glafka} and the references cited therein.
\bigbreak

The following references in relation to non-commutative calculus are useful in 
comparing with the present approach \cite{Connes, Dimakis, Forgy, MH}. Much of the present work is the fruit of a long
series of discussions with Pierre Noyes, Clive Kilmister and Anthony Deakin.  
The paper \cite{Mont} also works with minimal coupling for the Feynman-Dyson derivation. The first remark about
the minimal coupling occurs in the original paper by Dyson \cite{Dyson}, in the context of Poisson brackets.
The paper \cite{Hughes} is worth reading as a companion to Dyson.   It is the purpose of this summary to indicate how
non-commutative calculus can be used in foundations.
\bigbreak

\section{Constraints - Classical Physics and General Relativity}
The program here is to investigate restrictions in a non-commutative world that are imposed by asking for a specific correspondence between classical variables acting in the usual context of continuum calculus, and non-commutative operators corresponding to these classical variables. By asking for the simplest constraints we find the need for a quadratic Hamiltonian and a remarkable relationship with Einstein's equations for general relativity \cite{Deakin1,Deakin2}. There is a hierarchy of constraints of which we only analyze the first two levels. An appendix to this paper indicates a direction for exploring the algebra of the 
higher constraints.
\bigbreak

If, for example, we let $x$ and $y$ be classical variables and
$X$ and
$Y$ the corresponding  non-commutative operators, then we ask that $x^{n}$ correspond to $X^n$ and that $y^n$ correspond to $Y^n$ for
positive integers
$n.$ We further ask that linear combinations of classical variables correspond to linear combinations of the corresponding
operators. These restrictions tell us what happens to products. For example, we have classically that 
$(x+y)^2 = x^2 + 2xy + y^2.$ This, in turn must correspond to $(X + Y)^2 = X^2 + XY + YX + Y^2.$ From this it follows that
$2xy$ corresponds to $XY + YX.$ Hence $xy$ corresponds to 
$$\{ XY \} = (XY+YX)/2.$$
\bigbreak

By a similar calculation, if $x_1, x_2, \cdots , x_n$ are classical variables, then the product $x_{1}x_{2} \cdots x_{n}$ corresponds
to $$\{ X_{1}X_{2} \cdots X_{n} \} = (1/n!)\Sigma_{\sigma \in S_{n}} X_{\sigma_{1}}X_{\sigma_{2}} \cdots X_{\sigma_{n}}.$$
where $S_{n}$ denotes all permutations of $1,2,\cdots, n.$ Note that we use curly brackets for these symmetrizers and square 
brackets for commutators as in $[A,B]=AB-BA.$
\bigbreak

We can formulate constraints in the non-commutative world by asking for a correspondence between familiar differentiation formulas in
continuum calculus and the corresponding formulas in the non-commutative calculus, where all derivatives are expressed via commutators.
We will detail how this constraint algebra works in the first few cases. Exploration of these constraints has been 
pioneered by Anthony Deakin \cite{Deakin1,Deakin2,Deakin3}. The author of this paper and Tony Deakin have written a paper 
on the consequences of these contraints in the interface among classical and quantum mechanics and relativity \cite{KauffDeak}.
\bigbreak

Recall that the temporal derivative in a non-commutative world is represented by commutator with an operator $H$ that can be 
intrepreted as the Hamiltonian operator in certain contexts.
$$\dot{\Theta} = [\Theta,H].$$
For this discussion, we shall take a collection $Q^1,Q^2,\cdots,Q^n$ of operators to represent spatial coordinates
$q^1,q^2,\cdots,q^n$. The $Q^i$ commute with one another, and the derivatives with respect to $Q^i$ are represented by 
operators $P^i$ so that 
$$\partial{\Theta}/\partial{Q^i} = \Theta_i = [\Theta,P^i].$$
We also write $$\partial{\Theta}/\partial{P^i} = \Theta^i = [Q^i, \Theta].$$ 
Note that if $\Theta$ had indices of its own, then we would use a comma to separate indices indicating a derivative from the given indices. Thus
$$\partial{F_{a}}/\partial{Q^i} =  [F_{a}, Q^{i}] = F_{a,i}.$$ 
We assume that $[Q^i,P^j] = \delta^{ij}$ and that the $P^j$ commute with one another (so that mixed partial
derivatives with respect to the $Q^i$ are independent of order of differentiation).
\bigbreak

\noindent Note that $$\dot{Q^i} = [Q^i,H] = H^i.$$ It will be convenient for us to write
$H^i$ in place of  $\dot{Q^i}$ in the calculations to follow.
\bigbreak

\noindent {\bf The First Constraint.}  The {\it first constraint} is the equation
 $$\dot{\Theta} = \{\dot{Q^i} \Theta_i \} = \{ H^i \Theta_i \}.$$
This equation expresses the symmetrized version of the usual calculus formula
$\dot{\theta} = \dot{q^i}\theta_i .$ It is worth noting that the first constraint is satisfied by the 
quadratic Hamiltonian $$H = \frac{1}{4}(g^{ij}P^i P^j + P^i P^j g^{ij})$$ where $g^{ij} = g^{ji}$ and the $g_{ij}$ commute
with the $Q^k .$ One can show that a quadratic Hamiltonian is necessary for the first order constaint to be satisfied \cite{NCW,Glafka,Deakin2,KauffDeak}.
The fact that the quadratic Hamiltonian is equivalent to the first constraint shows how the constraints bind properties of classical physics (in this case Hamiltonian mechanics)
to the non-commutative world.
\bigbreak

\noindent {\bf The Second Constraint.} The {\it second constraint} is the symmetrized analog of the second temporal derivative:
$$\ddot{\Theta} = \{ \dot{H^i} \Theta_i \} + \{ H^i H^j \Theta_{ij} \}.$$
However, by differentiating the first constraint we have
$$\ddot{\Theta} = \{ \dot{H^i} \Theta_i \} + \{ H^i \{ H^j \Theta_{ij} \} \}$$
Thus the second constraint is equivalent to the equation 
$$\{ H^i \{ H^j \Theta_{ij} \} \}  = \{ H^i H^j \Theta_{ij} \}.$$
We now reformulate this version of the constraint in the following theorem.
\bigbreak

\noindent {\bf Theorem.} The second constraint in the form $\{ H^i \{ H^j \Theta_{ij} \} \}  = \{ H^i H^j \Theta_{ij} \}$
is equivalent to the equation
$$[[\Theta_{ij} , H^j], H^i] = 0.$$
\bigbreak

\noindent {\bf Proof.} We can shortcut the calculations involved in proving this Theorem by looking at the properties of 
symbols $A, B, C$ such that $AB = BA,$ $ACB = BCA.$ Formally these mimic the behaviour of $A = H^{i}, B= H^{j}, C = \Theta_{ij}$
in the expressions  $H^i H^j \Theta_{ij}$ and $H^i \Theta_{ij} H^j$ since $\Theta_{ij} = \Theta_{ji}$, and the Einstein summation
convention is in place. Then $$\{ A \{ BC \} \} = \frac{1}{4}(A(BC + CB) + (BC + CB)A)$$
$$ = \frac{1}{4}(ABC + ACB + BCA + CBA),$$
$$\{ ABC \} = \frac{1}{6}(ABC + ACB + BAC + BCA + CAB + CBA).$$
So $$\{ ABC \} - \{ A \{ BC \} \} = \frac{1}{12}(-ABC - ACB + 2BAC - BCA + 2CAB - CBA)$$
$$= \frac{1}{12}(ABC - 2ACB + CAB)$$
$$= \frac{1}{12}(ABC - 2BCA + CBA) $$
$$= \frac{1}{12}(A(BC - CB) + (CB - BC)A)$$
$$= \frac{1}{12}(A[B,C] - [B,C]A)$$
$$= \frac{1}{12}[A, [B,C]].$$
Thus the second constraint is equivalent to the equation
$$[ H^i ,[H^j , \Theta_{ij} ]] = 0.$$ This in turn is equivalent to the equation
$$[[\Theta_{ij} , H^j], H^i] = 0,$$ completing the proof of the Theorem.  \hfill\(\Box\) \\

\noindent {\bf Remark.}
If we define
$$\nabla^i(\Theta) = [\Theta, H^i] = [\Theta, \dot{Q^i}]$$
then this is the natural covariant derivative that was described in our discussion of non-commutative electromagnetism in Section 3 of this paper.
Thus the second order constraint is
$$\nabla^i (\nabla^j ( \Theta_{ij}) = 0.$$
\bigbreak

\noindent {\bf A Relationship with General Relativity.} 
\noindent We choose a non-commutative metric representative $g^{ij}$ in the non-commutative world with an inverse $g_{ij}$ so that $g^{ij} = g^{ji}, g_{ij} = g_{ji},$ and $g^{ik}g_{kj} = \delta^{i}_{j}.$
We can use the quadratic Hamiltonian $H = \frac{1}{4}(g^{ij}P^i P^j + P^i P^j g^{ij})$ as previously discussed, but we simplify the calculations below by taking $H= \frac{1}{2}(g^{ij}P^i P^j) .$
No essential difference ensues in the results.
We assume that the $g^{ij}$ commute with the coordinate representatives
$Q^{k}$ so that $[g^{ij}, Q^{k}] = 0$ for all choices of $i,j,k$ and similarly for the $g_{ij}.$ We take $P^{i}$ and $Q^{j}$ as described at the beginning of this section.
It is then an easy calculation to verify that $$[Q^{i}, \dot{Q^{j}}] = g^{ij}.$$  More generally, we have the\\ 

\noindent{\bf Lemma.} $\nabla^i(\Theta) = [\Theta, \dot{Q^i}] = g^{ij} [\Theta, P^{j}] = g^{ij}\Theta_{j}$ for an arbitrary element $\Theta$ in the non-commutative world algebra that commutes with the $g^{ij}$.\\

\noindent{\bf Proof.} $$\nabla^i(\Theta) = [\Theta, \dot{Q^i}] = [\Theta, [Q^{i}, H]] =  [\Theta, [Q^{i}, \frac{1}{2}(g^{ab}P^a P^b)]]= \frac{1}{2}g^{ab}[\Theta, [Q^{i}, P^a P^b]].$$
Note that $$[Q^{i}, P^aP^b]= Q^{i}P^{a}P^{b} - P^{a}P^{b}Q^{i} =Q^{i}P^{a}P^{b} - P^{a}Q^{i}P^{b} + P^{a}Q^{i}P^{b} -P^{a}P^{b}Q^{i}$$
$$=[Q^i , P^a ]P^b + P^a [Q^i , P^b] = \delta^{ia}P^{b} + P^{a}\delta^{ib}.$$ Therefore
 $$\nabla^i(\Theta) = \frac{1}{2}g^{ab}[\Theta, \delta^{ia}P^{b} + P^{a}\delta^{ib}] = \frac{1}{2}g^{ib}[\Theta, P^{b}] +  \frac{1}{2}g^{ai}[\Theta, P^{a}] = g^{ij}\Theta_{j}.$$
 This completes the proof of the Lemma. \hfill\(\Box\) \\

As we have seen in this section, the second order constraint is
$$\nabla^i (\nabla^j ( \Theta_{ij}) = 0.$$
Using the explicit form of the covariant derivative derived in the previous paragraph, we have 
$$\nabla^i (\nabla^j( \Theta_{ij}) = \nabla^i (g^{jk}\Theta_{ijk}) = g^{il} (g^{jk}\Theta_{ijk})_{l}$$
With $\Theta = g_{ab}$ the second constraint becomes the equation $$g^{il}(g^{jk}g_{ab,ijk})_{l} = 0.$$  
We call this equation the {\it specialized second order constraint}.
Kilmister observed in correspondence with Deakin \cite{Kilmister} 
that this last equation is, at the pole of canonical coordinates, equivalent to a fourth order version  of Einstein's field equation for vacuum general relativity:
$$K_{ab} = g^{ef}(R_{ab;ef} + \frac{2}{3} R_{ae}R_{fb})=0$$
where  $a,b,e,f = 1,2, \cdots n$ and $R$ is the curvature tensor corresponding to the metric $g_{ab}.$
This equation has been studied by Deakin in \cite{Deakin1,Deakin2,Deakin3} and by Deakin and Kauffman in \cite{KauffDeak}.
It remains to be seen what the full consequences for general relativity are in relation to this formulation, and it remains to be seen what the further consequences of higher order constraints will be.
The algebra of the higher order constraints is under investigation at this time.\\

\section{The Kilmister Equation}
In this section we derive the Kilmister equation $$K_{ab} = g^{ef}(R_{ab;ef} + \frac{2}{3} R_{ae}R_{fb})=0$$
where  $a,b,e,f = 1,2, \cdots 4$ and $R$ is the curvature tensor corresponding to the metric $g_{ab}.$ The derivation is based on explicating these tensors at the origin (pole) of canonical geodesic coordinates for 
spacetime with respect to the given metric. See Eddington \cite{Eddington} (page 79) for a detailed explanation of canonical coordinates. We will show that Kilmister's equation is, at the pole, equivalent to the specialized second order constraint equation  $$g^{ef}(g^{cd}g_{ab,cde})_{f} = 0$$ as explained in Section 4 of this paper. This
is a remarkable coincidence of structure and suggests that the Kilmister equation should be investigated in the context of general relativity and cosmology. Deakin and Kauffman have begun this investigation in \cite{KauffDeak}. In this section we give a complete derivation of the Kilmister equation based on the symmetries of the curvature and connection tensors. More work is needed to understand the relationship between this 
derivation and the structure of the second order constraint as described in the previous section of this paper.\\

Calculus in this section is classical continuum calculus. We use the standard notation $$F, a = \partial F/\partial x^{a}$$ where $x$ denotes a point in $4$-dimensional spacetime with $x^{4}$ the temporal coordinate. We use $F;a$ for the corresponding covariant derivative, which will be made explicit in the calculations below.\\

In order to perform Kilmister's derivation, we need to recall properties of the canonical coordinates and the basic 
symmetries of the Riemann tensor. For the present section we will refer to formal properties of the Riemann tensor and Levi-Civita connection as we need them. See the Appendix (Section 7)  for more details or Dirac's book on general relativity \cite{Dirac} for specifics about these tensors.\\

Eddingtion observes that in geodesic coordinates for four dimensional spacetime we may assume that the components $\Gamma^{a}_{bc}$ of the Levi-Civita connection
$$\Gamma^{a}_{bc} = \frac{g^{ak}}{2}(g_{kb,c} + g_{kc,b} - g_{bc,k})$$
vanish at that pole. Note that, in general, $\Gamma^{a}_{bc} = \Gamma^{a}_{cb}.$
Eddington further observes that one can assume, without constraining the curvature tensor, that 
$$\Gamma^{a}_{bc,d} + \Gamma^{a}_{cd,b} + \Gamma^{a}_{db,c} = 0$$ at the pole. Since the general formula for the Riemann tensor is
$$R^{a}_{bcd} = \Gamma^{a}_{bd,c} - \Gamma^{a}_{bc,d} + \Gamma^{k}_{bd}\Gamma^{a}_{kc} + \Gamma^{k}_{bc}\Gamma^{a}_{kd},$$
We know that  {\it at the pole}
$$R^{a}_{bcd} = \Gamma^{a}_{bd,c} - \Gamma^{a}_{bc,d}.$$

The general symmetries of the Riemann tensor that we use are:
\begin{enumerate}
\item $R_{abcd} = R_{cdab} = R_{dcba},$
\item$R_{abcd} = -R_{bacd} = -R_{abdc}.$
\end{enumerate}

\noindent {\bf Lemma.} At the pole of the canonical coordinates,  $\Gamma^{a}_{bc,d} = \frac{1}{3}(R^{a}_{bdc} + R^{a}_{cdb}).$\\

\noindent{\bf Proof.} At the pole, $$\Gamma^{a}_{bc,d} + \Gamma^{a}_{cd,b} + \Gamma^{a}_{db,c}  = 0$$ and at the pole
$$R^{a}_{bcd} = \Gamma^{a}_{bd,c} - \Gamma^{a}_{bc, d}.$$
Thus $$R^{a}_{bcd} = \Gamma^{a}_{bd,c} +\Gamma^{a}_{cd,b} + \Gamma^{a}_{db,c} = 2\Gamma^{a}_{bd,c} +\Gamma^{a}_{cd,b}.$$
Hence we have 
$$R^{a}_{bcd} = 2\Gamma^{a}_{bd,c} +\Gamma^{a}_{cd,b}$$ and
$$R^{a}_{cbd} = 2\Gamma^{a}_{cd,b} +\Gamma^{a}_{bd,c}.$$
Therefore
$$2R^{a}_{cbd} - R^{a}_{bcd} =  4\Gamma^{a}_{cd,b} + 2\Gamma^{a}_{bd,c} - 2\Gamma^{a}_{bd,c} - \Gamma^{a}_{cd,b} = 3 \Gamma^{a}_{cd,b}.$$
However, at the pole (and more generally), 
$$R^{a}_{bcd} + R^{a}_{cdb}+ R^{a}_{dbc} = \Gamma^{a}_{bd,c} - \Gamma^{a}_{bc, d} + \Gamma^{a}_{cb,d} - \Gamma^{a}_{cd,b} + \Gamma^{a}_{dc,b} - \Gamma^{a}_{db,c} = 0.$$
Therefore
$$3\Gamma^{a}_{cd, b} =  2R^{a}_{cbd} - R^{a}_{bcd}$$
$$= 2R^{a}_{cbd} + R^{a}_{cdb} + R^{a}_{dbc}$$
$$= R^{a}_{cbd} + R^{a}_{dbc}$$
since $R^{a}_{cbd} + R^{a}_{cdb} = 0$ by anti-symmetry in the indices $b$ and $d.$
Thus we have shown that 
$$\Gamma^{a}_{bc,d} = \frac{1}{3}(R^{a}_{bdc} + R^{a}_{cdb}).$$
This completes the proof of the Lemma. \hfill\(\Box\) \\

\noindent {\bf Lemma.} At the pole of the canonical coordinates, $$g_{ab,cd} = \frac{1}{3}(R_{cbad} + R_{cabd}).$$

\noindent{\bf Proof.} It is generally true that $$g_{ab,c} = g_{pb}\Gamma^{p}_{ac} + g_{ap}\Gamma^{p}_{bc}.$$
Thus, at the pole, $$g_{ab,cd} = g_{pb}\Gamma^{p}_{ac,d} + g_{ap}\Gamma^{p}_{bc,d}$$
$$=  g_{pb}[\frac{1}{3}(R^{p}_{adc} + R^{p}_{cda})] + g_{ap}[\frac{1}{3}(R^{p}_{bdc} + R^{p}_{cdb})]$$
$$= \frac{1}{3}(R_{badc} + R_{bcda} + R_{abdc} + R_{acdb})$$
$$ = \frac{1}{3}(R_{cbad} + R_{cabd})$$ (using the symmetries of the Riemann tensor).
This completes the proof of the Lemma. \hfill\(\Box\) \\

\noindent {\bf Definition.} Recall the definition of the Ricci Tensor: $$R_{ab} = g^{ij}R_{iabj} = R^{j}_{abj}.$$
Note that $$R_{ab} = g^{cd}R_{cabd} = g^{cd}R_{dbac} = R_{ba},$$ proving the symmetry of the Ricci Tensor.\\

\noindent{\bf Remark.} Since it is generally true that $$g_{ab,c} = g_{pb}\Gamma^{p}_{ac} + g_{ap}\Gamma^{p}_{bc},$$
we know that at the pole $$g_{ab,c} = 0,$$ since the Christoffel symbols vanish at the pole. 
Note that it follows from this vanishing result that $$g^{ab}_{, c} = 0$$ at the pole.
Higher derivatives may not be zero, as in the above Lemma.\\

\noindent {\bf Lemma.} At the pole of the canonical coordinates, $$g^{cd}g_{ab,cd} = \frac{2}{3}R_{ab}.$$

\noindent {\bf Proof.} By the previous Lemma
$$g^{cd}g_{ab,cd} = \frac{1}{3}(g^{cd}R_{cbad} + g^{cd}R_{cabd}) = \frac{1}{3}(R_{ba} + R_{ab}) = \frac{2}{3} R_{ab}.$$
This completes the proof of the Lemma. \hfill\(\Box\) \\

Now we are ready to obtain the Kilmister equation. For this, we need to invoke the covariant derivative, $R_{ab ;e},$ designated by a semi-colon, not a comma, and the basic formula
$$R_{ab,e} = R_{ab; e} + \Gamma^{p}_{ae}R_{pb} + \Gamma^{p}_{be}R_{ap}.$$ From this it follows that at the pole,
$$R_{ab,ef} = (R_{ab; e} + \Gamma^{p}_{ae}R_{pb} + \Gamma^{p}_{be}R_{ap})_{,f}$$
$$= R_{ab; ef} + \Gamma^{p}_{ae,f}R_{pb} + \Gamma^{p}_{be,f}R_{ap}.$$ 
Note that the other terms in this covariant derivative involve the Christoffel symbols and these vanish at the pole.
Thus we have 
$$R_{ab,ef} = R_{ab; ef} + \Gamma^{p}_{ae,f}R_{pb} + \Gamma^{p}_{be,f}R_{ap}$$ 
$$= R_{ab; ef} + \frac{1}{3}( R^{p}_{afe} R_{pb}  + R^{p}_{efa} R_{pb} + R^{p}_{bfe}R_{ap} + R^{p}_{efb}R_{ap} ).$$ 
Hence
$$g^{ef}R_{ab,ef} = g^{ef}R_{ab; ef} + \frac{1}{3}( g^{ef}R^{p}_{afe} R_{pb}  + g^{ef}R^{p}_{efa} R_{pb} + g^{ef}R^{p}_{bfe}R_{ap} + g^{ef}R^{p}_{efb}R_{ap} )$$
$$= g^{ef}R_{ab; ef} + \frac{1}{3}( 0  + g^{ef}R^{p}_{efa} R_{pb} + 0 + g^{ef}R^{p}_{efb}R_{ap} )$$
$$= g^{ef}R_{ab; ef} + \frac{1}{3}( R^{p}_{a} R_{pb} +  R^{p}_{b}R_{ap} )$$
$$= g^{ef}R_{ab; ef} + \frac{1}{3}( g^{ef}R_{ae} R_{fb} +  g^{ef}R_{eb}R_{af} )$$
$$= g^{ef}R_{ab; ef} + \frac{2}{3}( g^{ef}R_{ae} R_{fb}).$$
Thus we have shown 

\noindent {\bf Theorem.} At the pole of the canonical coordinates, $$g^{ef}R_{ab,ef} = K_{ab} = g^{ef}(R_{ab; ef} + \frac{2}{3}R_{ae} R_{fb})$$ and
$$g^{ef}R_{ab,ef} = \frac{3}{2} g^{ef}(g^{cd}g_{ab, cd})_{ef}.$$ 

\noindent{\bf Proof.} This result follows from the discussion above and the fact at the pole of the canonical coordinates, $$g^{cd}g_{ab,cd} = \frac{2}{3}R_{ab}.$$  \hfill\(\Box\) \\

\noindent {\bf Remark.} Thus we have shown that $$\frac{2}{3}K_{ab} = g^{ef}(g^{cd}g_{ab, cd})_{ef}$$
$$ =  g^{ef}(g^{cd}_{,e}g_{ab, cd})_{f} + g^{ef}(g^{cd}g_{ab, cde})_{f} = g^{ef}(g^{cd}g_{ab, cde})_{f},$$
since $g^{cd}_{,e} = 0.$  Thus we have proved:\\

\noindent {\bf Theorem. } Let $$K_{ab} = g^{ef}(R_{ab; ef} + \frac{2}{3}R_{ae} R_{fb}).$$ Then
$$\frac{2}{3}K_{ab} = g^{ef}(g^{cd}g_{ab,cde})_{f}.$$ Thus the Kilmister equation $K_{ab} = 0$ is equivalent to the second order constraint articulated in Section 4 of the present paper.  \\

This completes our description of Clive Kilmister's remarkable derivation of the relationship of the second order constraint with general relativity.
All these considerations are motivation for considering the Kilmister tensor equation $K_{ab}= 0$ as a refined version of the vacuum equations for general relativity.
In \cite{KauffDeak} we explore some of the consequences of the Kilmister equation. The exact relationship of the constraint equation and the Kilmister equation remains mysterious. More work needs to be done 
in this domain and in exploring the relationship of non-commutative worlds and the tensor geometry of classical spacetime.\\
\\

\section {\bf On the Algebra of Constraints}
 We have the usual advanced calculus formula
 $\dot{\theta} = \dot{q^i}\theta_i .$
 We shall define $h^{j} =  \dot{q^i}$ so that we can write $\dot{\theta} = h^{i}\theta_i .$
 We can then calculate successive derivatives with $\theta^{(n)}$ denoting the $n$-th temporal derivative of $\theta.$
 $$\theta^{(1)} = h^{i}\theta_i $$
 $$\theta^{(2)} = h^{i(1)}\theta_i + h^{i}h^{j}\theta_{ij}$$ 
 $$\theta^{(3)} = h^{i(2)}\theta_i + 3h^{i(1)}h^{j}\theta_{ij}  + h^{i}h^{j}h^{k}\theta_{ijk}$$ 
The equality of mixed partial derivatives in these calculations makes it evident that one can use a formalism that hides all the superscripts and subscripts ($i,j,k, \cdots)$). In that simplified formalism, we can write
 $$\theta^{(1)} = h\theta$$
 $$\theta^{(2)} = h^{(1)}\theta + h^{2}\theta$$ 
 $$\theta^{(3)} = h^{(2)}\theta + 3h^{(1)}h \theta  + h^{3} \theta$$ 
 $$\theta^{(4)} =h^4 \theta+6 h^2 \theta h^{(1)}+3 \theta h^{(1)2}+4 h\theta h^{(2)}+\theta h^{(3)}$$
Each successive row is obained from the previous row by applying the identity $\theta^{(1)} = h\theta$
in conjunction with the product rule for the derivative. 
\bigbreak

This procedure can be automated so that one can obtain the formulas for higher order derivatives as far as one desires. These can then be converted into the non-commutative constraint algebra and the 
consequences examined. Further analysis of this kind will be done in a sequel to this paper.
\bigbreak

The interested reader may enjoy seeing how this formalism can be carried out.
Below we illustrate a calculation using $Mathematica^{TM},$ where the program already knows how to 
formally differentiate using the product rule and so only needs to be told that $\theta^{(1)} = h\theta.$
This is said in the equation $T'[x] = H[x] T[x]$ where $T[x]$ stands of $\theta$ and $H[x]$ stands
for $h$ with $x$ a dummy variable for the differentiation. Here $D[T[x],x]$ denotes the derivative of 
$T[x]$ with respect to $x$, as does $T'[x],$
\bigbreak

In the calculation below we have indicated five levels of derivative. The structure of the coefficients in this recursion is interesting and complex territory.
For example, the coefficients of $H[x]^n T[x] H'[x] = h^n \theta h'$ are the triangular numbers 
$\Set{1,3,6,10,15,21, \cdots }$ but the next series are the coefficients
of $H[x]^n T[x] H'[x]^2 = h^n \theta h'^{2},$ and these form the series 
$$\Set{1,3,15,45,105,210,378,630,990,1485,2145, \cdots }.$$ This series is eventually constant after four discrete differentiations. This is the next
simplest series that occurs in this structure after the triangular numbers. To penetrate the full algebra of constraints we need to understand the structure
of these derivatives and their corresponding non-commutative symmetrizations.
\bigbreak

\noindent\(\pmb{}\)

\noindent\(\pmb{}\\
\pmb{T'[x] \text{:=} H[x] T[x]}\)

\noindent\(\pmb{}\\
\pmb{D[T[x],x]}\\
\pmb{D[D[T[x],x],x]}\\
\pmb{D[D[D[T[x],x],x],x]}\\
\pmb{D[D[D[D[T[x],x],x],x],x]}\\
\pmb{D[D[D[D[D[T[x],x],x],x],x],x]}\)
 
\noindent\(H[x] T[x]\)
\bigbreak

\noindent\(H[x]^2 T[x]+T[x] H'[x]\)
\bigbreak

\noindent\(H[x]^3 T[x]+3 H[x] T[x] H'[x]+T[x] H''[x]\)
\bigbreak

\noindent\(H[x]^4 T[x]+6 H[x]^2 T[x] H'[x]+3 T[x] H'[x]^2+4 H[x] T[x] H''[x]+T[x] H^{(3)}[x]\)
\bigbreak

\noindent\(H[x]^5 T[x]+10 H[x]^3 T[x] H'[x]+15 H[x] T[x] H'[x]^2+10 H[x]^2 T[x] H''[x]+10 T[x] H'[x] H''[x]+5 H[x] T[x] H^{(3)}[x]+T[x] H^{(4)}[x]\)
\bigbreak

\subsection{Algebra of Constraints}
In this section we work with the hidden index conventions described before in the paper.
In this form, the classical versions of the first two constraint equations are
\begin{enumerate}
\item $\dot{\theta} = \theta h$
\item $\ddot{\theta} = \theta h^{2} + \theta \dot{h}$
\end{enumerate}

In order to obtain the non-commutative versions of these equations, we replace $h$ by $H$ and
$\theta$ by $\Theta$ where the capitalized versions are non-commuting operators. The first and second constraints then become
\begin{enumerate}
\item $\{ \dot{\Theta} \} = \{ \Theta H \} = \frac{1}{2}(\Theta H + H \Theta)$
\item$\{ \ddot{\Theta} \} = \{ \Theta H^{2} \} + \{ \Theta \dot{H} \} = 
\frac{1}{3} (\Theta H^{2} + H \Theta H + H^{2} \Theta) + \frac{1}{2}(\Theta \dot{H} + \dot{H} \Theta)$
\end{enumerate}

\noindent {\bf Proposition.} The Second Constraint is equivalent to the commutator equation
$$[[\Theta, H], H] = 0.$$
\bigbreak

\noindent {\bf Proof.} We identify 
$$ \{ \dot{\Theta} \}^{\bullet} = \{ \ddot{\Theta} \}$$
and
$$ \{ \dot{\Theta} \}^{\bullet} = \{ \{ \Theta H \} H \} + \{ \Theta \dot{H} \}.$$
So we need
$$\{ \Theta H^{2} \} = \{ \{ \Theta H \} H \}.$$
The explicit formula for $ \{ \{ \Theta H \} H \}$ is
$$\{ \{ \Theta H \} H \} = \frac{1}{2}(\{ \Theta H \} H + H \{ \Theta H \} ) = \frac{1}{4}(\theta H H + H \Theta H + H \Theta H + H H \Theta ).$$
Thus we require that 
$$\frac{1}{3} (\Theta H^{2} + H \Theta H + H^{2} \Theta) = \frac{1}{4}(\theta H H + H \Theta H + H \Theta H + H H \Theta ).$$
which is equivalent to 
$$\Theta H^{2} + H^{2} \Theta - 2 H \Theta H = 0.$$
We then note that 
$$[[\Theta, H], H] = (\Theta H - H \Theta) H - H (\Theta H - H \Theta) = \Theta H^{2} + H^{2} \Theta - 2 H \Theta H.$$
Thus the final form of the second constraint is the equation
$$[[\Theta, H], H] = 0. $$  \hfill\(\Box\) \\

\noindent {\bf The Third Constraint.} We now go on to an analysis of the third constraint.
The third constraint consists in the the two equations
\begin{enumerate}
\item $\Set{\dddot{\Theta}} = \Set{\Theta H^3 } + 3\Set{\Theta H \dot{H}} + \Set{\Theta \ddot{H}}$
\item $\Set{\dddot{\Theta}} = \Set{\ddot{\Theta}}^{\bullet}$ where
$$ \Set{ \ddot{\Theta}}^{\bullet}  = \Set{ \Set{\Theta H} H^2  } + 2 \Set{ \Theta H \dot{H} } + \Set{\Set{\Theta H} \dot{H}} + \Set{\Theta \ddot{H}} $$
\end{enumerate}
\bigbreak

\noindent {\bf Proposition.} The Third Constraint is equivalent to the commutator equation
$$[H^2 , [H, \Theta]] = [\dot{H}, [H, \Theta]] - 2[H, [\dot{H}, \Theta]].$$ 
\bigbreak

\noindent {\bf Proof.} 
We demand that $\Set{\dddot{\Theta}} = \Set{\ddot{\Theta}}^{\bullet}$ and this becomes the longer equation
$$\Set{\Theta H^3 } + 3\Set{\Theta H \dot{H}} + \Set{\Theta \ddot{H}} = \Set{ \Set{\Theta H} H^2  } + 2 \Set{ \Theta H \dot{H} } + \Set{\Set{\Theta H} \dot{H}} + \Set{\Theta \ddot{H}} $$
This is equivalent to the equation
$$\Set{\Theta H^3 } + \Set{\Theta H \dot{H}}  = \Set{ \Set{\Theta H} H^2  } + \Set{\Set{\Theta H} \dot{H}} $$
This, in turn is equivalent to 
$$\Set{\Theta H^3 }   - \Set{ \Set{\Theta H} H^2  } =  \Set{\Set{\Theta H} \dot{H}} -  \Set{\Theta H \dot{H}}$$
This is equivalent to 
$$(1/4)(H^3 \Theta + H^2 \Theta H + H \Theta H^2 + \Theta H^3) - (1/6)(H^2 (H \Theta + \Theta H) + H (H \Theta + \Theta H) H + (H \Theta + \Theta H) H^2) $$ 
$$= (1/2)(\dot{H} (1/2)(H \Theta + \Theta H)  + (1/2)(H \Theta + \Theta H) \dot{H}) - (1/6)(\dot{H} H \Theta + \dot{H} \Theta H +  H \dot{H} \Theta +  H \Theta \dot{H} +  \Theta H \dot{H} + \Theta  \dot{H} H)$$
This is equivalent to 
$$3(H^3 \Theta + H^2 \Theta H + H \Theta H^2 + \Theta H^3) - 2(H^3 \Theta + 2 H^2 \Theta H  + 2 H \Theta H^2 + \Theta H^3) $$ 
$$= 3(\dot{H} H \Theta + \dot{H} \Theta H  + H \Theta \dot{H} + \Theta H \dot{H}) - 2(\dot{H} H \Theta + \dot{H} \Theta H +  H \dot{H} \Theta +  H \Theta \dot{H} +  \Theta H \dot{H} + \Theta  \dot{H} H)$$
This is equivalent to 
$$H^3 \Theta - H^2 \Theta H - H \Theta H^2 + \Theta H^3 $$ 
$$= (\dot{H} H \Theta + \dot{H} \Theta H  + H \Theta \dot{H} + \Theta H \dot{H}) - 
2( H \dot{H} \Theta + \Theta  \dot{H} H)$$
The reader can now easily verify that 
$$[H^2 , [H, \Theta]] = H^3 \Theta - H^2 \Theta H - H \Theta H^2 + \Theta H^3 $$
and that 
$$[\dot{H}, [H, \Theta]] - 2[H, [\dot{H}, \Theta]] = (\dot{H} H \Theta + \dot{H} \Theta H  + H \Theta \dot{H} + \Theta H \dot{H}) - 2( H \dot{H} \Theta + \Theta  \dot{H} H)$$
Thus we have proved that the third constraint equations are equivalent to the commutator equation
$$[H^2 , [H, \Theta]] = [\dot{H}, [H, \Theta]] - 2[H, [\dot{H}, \Theta]]$$ 
This completes the proof of the Proposition.  \hfill\(\Box\) \\

\noindent {\bf Discussion.} Each successive constraint involves the explicit formula for the 
higher derivatives of $\Theta$ coupled with the extra constraint that 
$$\Set{ \Theta^{(n)} }^{\bullet} = \Set{  \Theta^{(n+1)} }.$$ We conjecture that each constraint can be expressed as a commutator equation in terms of $\Theta$ , $H$ and the derivatives of $H,$ in analogy to the formulas that we have found for the first three constraints. This project will continue with a deeper algebraic study of the constraints and their physical meanings.
\bigbreak

\section{Appendix -- Einstein's Equations and the Bianchi Identity}
The purpose of this section is to show how the Bianchi identity (see below for its definition) appears in the context of 
non-commutative worlds. The Bianchi identity is a crucial mathematical ingredient in general relativity. We shall begin
with a quick review of the mathematical structure of general relativity (see for example \cite{FN}) and then turn to the context of
non-commutative  worlds.
\bigbreak

The basic tensor in Einstein's theory of general relativity is $$G^{ab} = R^{ab} - \frac{1}{2}Rg^{ab}$$ where
$R^{ab}$  is the Ricci tensor and $R$ the scalar curvature. The Ricci tensor and the scalar curvature are both obtained
by contraction from the Riemann curvature tensor $R^{a}_{bcd}$ with $R_{ab} = R^{c}_{abc}, R^{ab} = g^{ai}g^{bj}R_{ij},$ and 
$R = g^{ij}R_{ij}.$ Because the Einstein tensor $G^{ab}$ has vanishing divergence, it is a prime candidate to be proportional to the
energy momentum tensor $T^{\mu \nu}.$ The Einstein field equations are $$R^{\mu \nu} - \frac{1}{2}Rg^{\mu \nu} = \kappa T^{\mu \nu}.$$
\bigbreak

The reader may wish to recall that the Riemann tensor is obtained from the commutator of a covariant derivative $\nabla_k,$
associated with the Levi-Civita connection $\Gamma^{i}_{jk} = (\Gamma_{k})^{i}_{j}$ (built from the space-time metric $g_{ij}$).
One has  
$$\lambda_{a:b} = \nabla_{b}\lambda_{a} = \partial_{b}\lambda_{a} - \Gamma^{d}_{ab}\lambda_{d}$$ or
$$\lambda_{:b}= \nabla_{b}\lambda = \partial_{b}\lambda - \Gamma_{b}\lambda$$ for a vector field $\lambda.$ 
With $$R_{ij} = [\nabla_{i}, \nabla_{j}] = \partial_{j}\Gamma_{i} - \partial_{i}\Gamma_{j} + [\Gamma_{i}, \Gamma_{j}],$$ one has
$$R^{a}_{bcd} = (R_{cd})^{a}_{b}.$$ (Here $R_{cd}$ is {\it not} the Ricci tensor. It is the Riemann tensor with two internal indices
hidden from sight.)

One way to understand the mathematical source of the Einstein tensor, and the vanishing of its divergence, is to see it as a
contraction of the Bianchi identity for the Riemann tensor. The Bianchi identity states
$$R^{a}_{bcd:e} + R^{a}_{bde:c} + R^{a}_{bec:d} = 0$$ where the index after the colon indicates the covariant derivative.
Note also that this can be written in the form 
$$(R_{cd:e})^{a}_{b} + (R_{de:c})^{a}_{b} + (R_{ec:d})^{a}_{b} = 0.$$
The Bianchi identity is a consequence of local properties of the Levi-Civita connection and  consequent symmetries of the Riemann
tensor. One relevant symmetry of the Riemann tensor is the equation
$R^{a}_{bcd} = - R^{a}_{bdc}.$

We will not give a classical derivation of the Bianchi identity here, but it is
instructive to see how its contraction leads to the Einstein tensor. To this end, note that we can contract the Bianchi identity to
$$R^{a}_{bca:e} + R^{a}_{bae:c} + R^{a}_{bec:a} = 0$$ which, in the light of the above definition of the Ricci tensor and the 
symmetries of the Riemann tensor  is the same as
$$R_{bc:e} - R_{be:c} + R^{a}_{bec:a} = 0.$$ Contract this tensor equation once more to obtain
$$R_{bc:b} - R_{bb:c} + R^{a}_{bbc:a} = 0,$$ and raise indices
$$R^{b}_{c:b} - R_{:c} + R^{ab}_{bc:a} = 0.$$ Further symmetry gives
$$R^{ab}_{bc:a} = R^{ba}_{cb:a} = R^{a}_{c:a} = R^{b}_{c:b}.$$ Hence we have
$$2R^{b}_{c:b} - R_{:c} = 0,$$ which is equivalent to the equation
$$(R^{b}_{c} - \frac{1}{2}R\delta^{b}_{c})_{:b} = G^{b}_{c:b} = 0.$$ From this we conclude
that $G^{bc}_{:b} = 0.$ The Einstein tensor has appeared on the stage with vanishing divergence, courtesy of the Bianchi 
identity!
\bigbreak

\noindent {\bf Bianchi Identity and Jacobi Identity.} Now lets turn to the context of non-commutative worlds. We
have infinitely many  possible convariant derivatives, all of the form $$F_{:a} = \nabla_{a}F = [F, N_{a}]$$ for some $N_{a}$ elements
in the non-commutative  world. Choose any such covariant derivative. Then, as in the introduction to this paper, we have the curvature
$$R_{ij} = [N_{i},N_{j}]$$ that represents the commutator of the covariant derivative with itself in the sense that 
$[\nabla_i,\nabla_j]F = [[N_i,N_j],F].$ Note that $R_{ij}$ is not a Ricci tensor, but rather the indication of the external structure
of the curvature without any particular choice of linear representation (as is given in the classical case as described above).
We then have the Jacobi identity
$$[[N_a,N_b],N_c] + [[N_c,N_a],N_b] + [[N_b,N_c],N_a] = 0.$$ Writing the Jacobi identity in terms of curvature and covariant
differention we have $$R_{ab:c} + R_{ca:b} + R_{bc:a}.$$ Thus in a non-commutative world, every covariant derivative
satisfies its own Bianchi identity. This gives an impetus to study general relativity in non-commutative worlds by 
looking for covariant derivatives that satisfy the symmetries of the Riemann tensor and link with a metric in an appropriate way.
We have only begun this aspect of the investigation. The point of this section has been to show the intimate relationship between the
Bianchi idenity and the Jacobi identity that is revealed in the context of non-commutative worlds.
\bigbreak

\end{document}